\documentclass[aps,preprint,twocolumn]{revtex4}
\usepackage{epsfig}
\usepackage{amsmath}
\usepackage{epsfig}
\usepackage{amssymb}

\begin{document}
\title{An interesting track for the Brachistochrone}
\author{Zafar Ahmed$^{1,3}$ and Amal Nathan Joseph$^{2,3}$}
\affiliation{$~^1$Nuclear Physics Division,$~^2$Reactor Physics Design Division, Bhabha Atomic, Research Centre, Trombay, Mumbai, 400085 \\
$~^3$Homi Bhabha National Institute, Anushaktinagar, Mumbai, 400094}
\date{\today}
\begin{abstract}
If a particle  has to fall first vertically 1 m  from A and then move horizontally 1 m to B, it takes a time $t(=\tau_1+\tau_2=\tau_3=\frac{3}{\sqrt{2g}})=0.67$ s. Under gravity and without friction, if it sides   down on a linear track inclined at $45^0$ between two points A and B of 1 m height, it takes time $t(=\tau_4=\frac{2}{\sqrt{g}})=0.63$ s. Between these two extremes, historically, Bernoulli (1718) proved that  the fastest track between these points A and B is cycloid  with the least time of descent  $t=\tau_B=0.58$ s. Apart from other interesting cases, here we study the frictionless motion of a particle/bead on an interesting track/wire between A and B  given by $y(x)=(1-x^{\nu})^{1/\nu}.$ For $\nu > 1$ the track becomes convex and $t>>\tau_4$, and when $\nu >1.22$, the motion with zero initial speed is not possible. We find that when $\nu \in (0.09653, 0.31749), \tau_4<t <\tau_3$ and when $\nu \in ( 0.31749, 1),\tau_B < t < \tau_4$. But most remarkably, the concave curve becomes very steep/deep if $\nu \in (0, \nu_c=0.09653)$, then $t=0.2258$ s $< \tau_B$, this is as though  a particle would travel 1 meter horizontally with a speed equal $\sqrt{2g}$ m/sec to take the time ($=\frac{1}{\sqrt{2g}}=\tau_2) < \tau_B$. The function $t(\nu$) suffers a jump discontinuity at $\nu=\nu_c$, we offer some resolution.
\end{abstract}
\maketitle
Ignoring  friction and taking the acceleration due to gravity $g=9.8$ m/s, let us first appreciate the various times a particle would take in the following motions under  gravity: (1) in free falling from a height of 1 m, $t=\tau_{1}=\sqrt{\frac{2}{g}}=0.4517$ s. (2) in moving a distance of 1 m after falling, $t=\tau_2=\sqrt{\frac{1}{2g}}=0.2258$ s. Let $\tau_3=\tau_{1}+\tau_2=\frac{3}{\sqrt{2g}}=0.6776$ s. (3)  in falling 1 meter on an inclined straight line (Fig.1) at $45^0$, $t=\tau_4=\frac{2}{\sqrt{g}}=0.6338$ s. Bernoulli (1718) found that the classical Brachistochrone (least time) problem gave the cycloid (like {\bf b} in Fig.1) as the fastest descent curve between two points  and the time $t=\tau_B=\frac{1}{\sqrt{2g}}\int_{0}^{\pi/2}\sqrt{\csc x} dx=0.5822$s [1] which is less than $\tau_4.$ The distance between the end points of this cycloid is $AB=\sqrt{2}$ m.

\begin{figure}[t]
	\centering
	\includegraphics[width=7cm,height=5 cm,scale=1.]{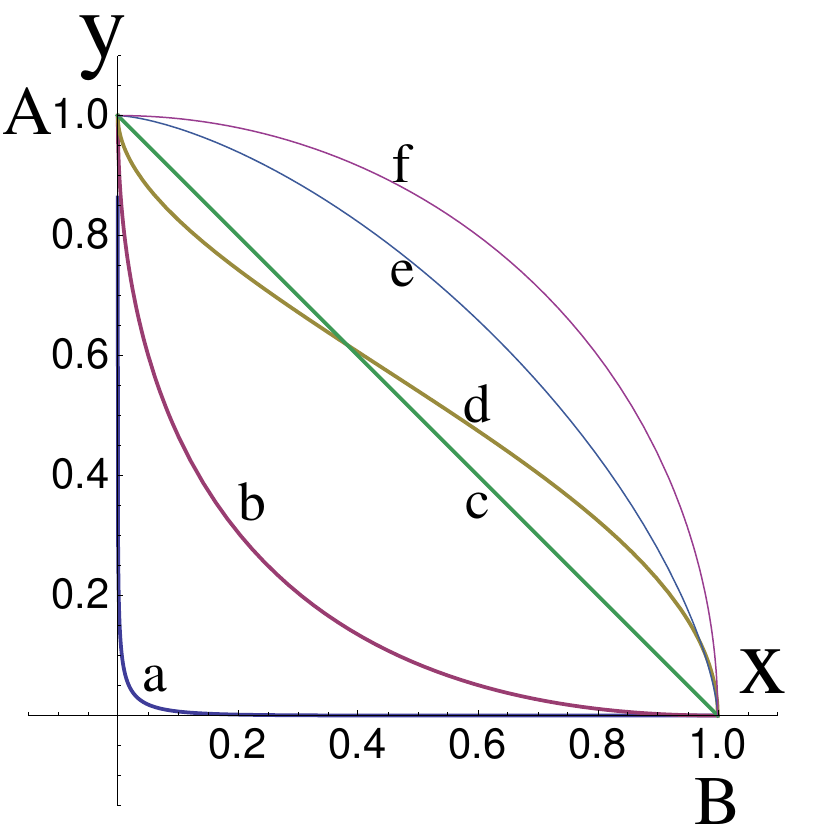}
	\caption{For various values of $\nu$ the tracks given by (7). a: $\nu=0.2, b: \nu=0.5, d: \nu=1, e: \nu=1.2, f: \nu=2.$ The track d denotes one of the family of $\mu$-curves given by (7). These $\mu$-tracks are essentially vertical at both points A and B.}
\end{figure}

The sliding of a small bead under gravity on an inclined and curved (Fig.1) tracks (wires) is what makes most simple model for Brachistochrone. This problem is well discussed in the textbooks  [2] using the simple equation of the energy conservation and the optimization by variational principle. 
Terrestrial [3]  Brachistochrone and several variants consisting in central force field [4], bent straight wire [5] and unrestrained rolling block [6] have been discussed. A simplified method for the direct and inverse problem of finding the force given the  Brachistochrone curve has been developed [7]. Minimum descent time  along a set of connected  inclined planes has also been explored [8]. A discrete Brachistochrone  has been proposed [9]. Quantum Brachistochrone problem has also been defined and formulated in terms of Hermitian, complex PT-symmmetric [10]  and pseudo-Hermitian [11] Hamiltonians.

The question of other types of interesting tracks (e.g., Fig.1) and their features are  discussed here.  We study the time of descent from  various tracks given by $y(x)=1-x^{\lambda}$, $y(x)=\sqrt{1-x^\mu}$ and $y(x)=(1-x^\nu)^{1/\nu}$ between two fixed points A(0,1) and B(1,0). It is the third $\nu$-track that gives a surprising result that yields $t$ lesser than the so far acclaimed least value $\tau_B$, when $\nu \in (0.0,0.09653)$. 

 When a particle of mass $m$ sides under gravity on the track given by $y(x)$, the conservation of energy leads to
 \begin{equation}
 \frac{m}{2}[\dot x^2+ \dot y^2]+mgy(x)=mgy(x_1) 
 \end{equation}
 Introducing differential  length of the curve $(ds)^2=(dx)^2+(dy)^2$, we can write
 \begin{equation}
 \frac{dt}{dx}=\sqrt{\frac{1+y'^2(x)}{[2g(y(x_1)-y(x)]}}
 \end{equation}
 By taking the initial point as $x_1=0$ and the final point as $x_2=1$. Here the distances are in meters and  $g=9.8$ m/s,
 we get 
 \begin{equation}
 t=\tau_2 \int_{0}^{1} \sqrt{\frac{1+y'^2(x)}{y(0)-y(x)}}dx, ~ \tau_2=\frac{1}{\sqrt{2g}}
 \end{equation}
 For the linear track-{\bf c} (Fig.1) $y=a(1-x)$, where $a=\tan \theta >0$, we  get 
 \begin{equation}
 t=\frac{1}{\sqrt{g}} \sqrt{\frac{1+a^2}{2a}} \int_{0}^{1} \frac{dx}{\sqrt{x}} \ge  \frac{2}{\sqrt{g}}=\tau_4
 \end{equation}
  
 As $(1-a)^2\ge 0 \implies 1+a^2 \ge 2a$ and the equality holds for $a=1 \implies \theta=\pi/4$. This proves that among all the linear inclined tracks sarting from A and ending at $x=1$, the  one with $45^0$ angle gives the least time of descent, Galileo is known to have pointed out this fact.
 
 First, we study the curved track given by
 \begin{equation}
 y(x)=1-x^{\lambda}, \lambda>0.
 \end{equation}
 Using (3), we find that for this track 
  \begin{equation}
 t(\lambda)=\tau_2 \int_{0}^{1} \frac{\sqrt{1+4\lambda^2 x^{2\lambda-2}}}{x^{\lambda/2}} dx,
 \end{equation}
 $t(0)=\tau_2.$
 This is an improper integral as its integrand diverges at $x=0$, however, it is convergent (finite) when $\lambda <2$, like in Eq. (4). Eq. (6) can be written in terms of Gamma functions or Gauss hypergeometric function, nevertheless it is easily doable numerically. We have used ``NIntegrate" of Mathematica.
 
 \begin{figure}[t]
	\centering
	\includegraphics[width=7cm,height=5 cm,scale=1.0]{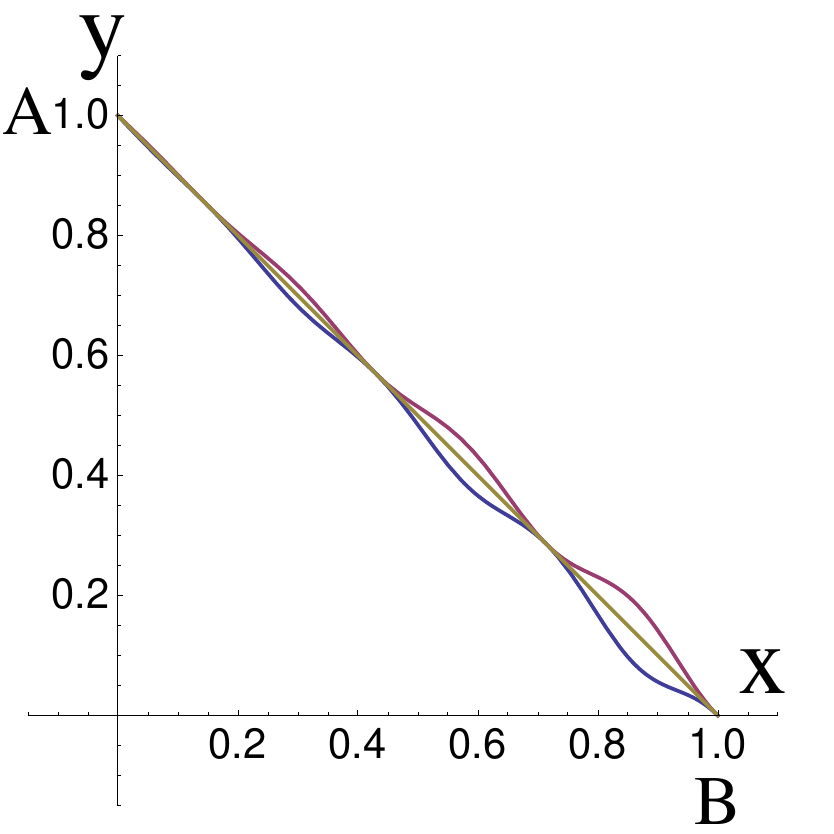}
	\caption{The wavy tracks $y(x)=1-xe^{\epsilon\cos^2(7\pi x/2)}$. The lower(upper) one is for $\epsilon=\pm0.06$. Both the tracks up to $x=0.2$ are almost linear, but after this, the lower one is concave, this gives the particle more speed to slide faster ($ t=0.6373$) than the upper one ($t=0.6478$). Interstingly the length of the upper track is a fraction less than the lower one.}
\end{figure}
 
     Fig.1, presents the $\lambda$-tracks only schematically. For $\lambda \in (0,0.07)$, see the  track-{\bf a} is concave but deep/steep, we get $\tau_4<t(\lambda)<\tau_3$. This parametric regime in a limiting way represents motion of the particle vertically down  from point A, followed by the horizontal motion with speed $\sqrt{2g}$ m/s towards the point B.  When $\lambda \in (.07, 1)$ the track-{\bf b} is concave, $\tau_B<t(\lambda) < \tau_4$ . Next, when $\lambda \in (1,2)$, see the convex track-{\bf e}, we find that $t(\lambda)>>\tau_4$. The motion is forbidden, when  $\lambda \ge 2$, $t(\lambda)$ (6) diverges: the convexity of the track-{\bf f}  does not allow the motion with zero initial velocity.

We can have  wavvy tracks (see Fig. 2) that are given by $y(x)=1-xf(x), 1-\sqrt{x} f(x), 1-x^{3/2} f(x)$.  Here we choose $f(x)=e^{\epsilon \cos^2(7\pi x/2)}, \epsilon=\pm 0.06$ to give a mild and smooth modulation to the tracks. These small modulations in  the case of  concave or convex tracks do not change the inequality of the time of descent  $t$, with respect to $\tau_4$. The essence of this study is that if initial part of a track is concave for a good length like the lower one, then $t=0.6373$ is lesser than for  the upper track $t=0.6478$, even though its length is a little less, but its initial part is convex. However, in both the cases $t>\tau_4$
  
The family of tracks given by
  \begin{equation}
  y(x)=\sqrt{1-x^\mu}, \mu >0
  \end{equation}
  are vertical at $x=0$ and $x=1$, see track-{\bf d} for $\mu=0.5$. We find that when $\mu \in(0,0.0161)$, $\tau_4<t(\mu) < \tau_3$, for $\mu \in (0.0161, 0.615),  t(\mu) <\tau_4$, for $\mu \in (0.615, 1.23)$, $t(\mu) >>\tau_4$. For $\mu \in (1.23, 1.61)$,  we get $t(\mu) >>\tau_3$ with a warning of accuracy. For $\mu>1.61$ the integral (4) starts diverging and the convexity of the track does not allow the motion with  zero initial speed. For $\mu=0.5$, see the track-{\bf d}, $t=0.6211<\tau_4$, here the initial part of the track is concave followed by a convex part.
  
  We would like to remark that for both $\lambda$ and $\mu$ tracks, the numerical integration yields: $\tau_B<t(0^+) \le \tau_3$  with only a warning of accuracy.
  
   More interestingly, the  tracks given by
   \begin{equation}
   y(x)=(1-x^\nu)^{1/\nu}, \nu >0
   \end{equation}
 From (3), we get  $t(\nu)=\tau_2 I(\nu)$ in terms of an interesting integral  
\begin{small}
 \begin{equation}
I(\nu){=} \int_{0}^{1}\sqrt{\frac{x^{2(\nu-1)}+(1-x^\nu)^{2(1-\nu)/\nu}}{x^{2(1-\nu)/\nu}[1-(1-x^{\nu})^{1/\nu}]}} dx
 \end{equation}
 \end{small}
 The simple interesting cases are $I(0)=1$,$ I(1)=2\sqrt{2}$ and 
 \begin{small}
\begin{equation}
I(\nu)= \int_{0} 
 \frac{dx}{x^{(3\nu-2)/2}}<\infty, ~ if~ \nu<4/3.
 \end{equation}
 \end{small}
 NIntegrate of Mathematica gives $I(0\le \nu<\nu_c)=1$ and $I(\nu_c<\nu<\nu_s)=3$
 without a warning  for any type of error and shows $I(\nu>\nu_s)$ as divergent, where $\nu_s=1.22$. $I(\nu)$ for $\nu \in (0,1.22)$  has a jump discontinuity at $\nu=\nu_c$, see Fig. 2.
 
 For $\nu > 1$ the track becomes convex and $t>>\tau_1$, and when $\nu >1.22$, the motion with zero initial speed is not possible as the integral (3) diverges. We find that when $\nu \in (0.09653, 0.31749), \tau_4<t <\tau_3$ and when $\nu \in ( 0.31749, 1),  \tau_B <t(\nu) < \tau_3$. But most remarkably, if $\nu \in (0, 0.09653=\nu_c)$, the track is concave but deep and steep  then $t=\tau_2=0.2258$ sec $< \tau_B$, this is the time ($=\frac{1}{\sqrt{2g}}=\tau_2)$ that a particle would  take to travel 1 meter horizontally with a speed equal $\sqrt{2g}$ m/sec which is acquired at the origin. The function $t(\nu$) suffers a jump discontinuity at $\nu=\nu_c$, see Fig. 3.  
  \begin{figure}
	\centering
	\includegraphics[width=7cm,height=5 cm,scale=1.0]{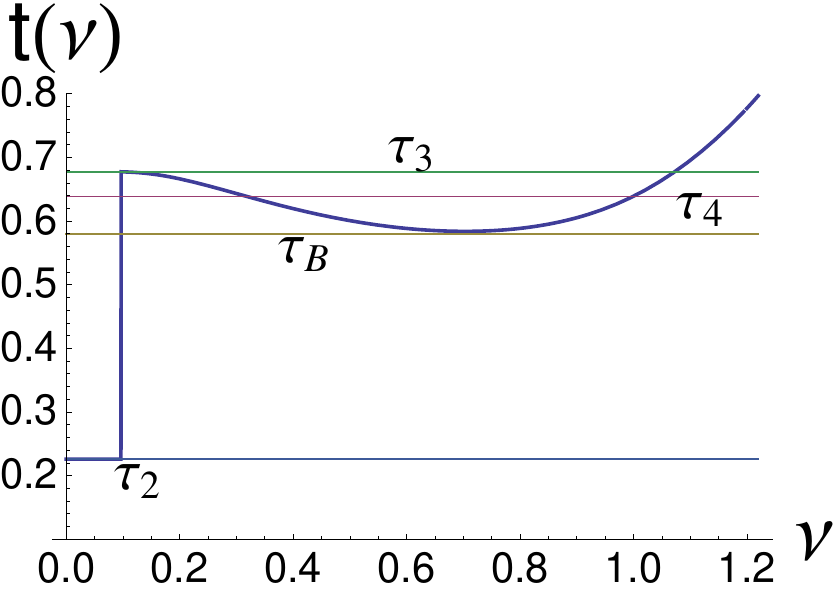}
	\caption{ $t(\nu)$ for the track (8). In the domains  $\nu=(0.09653,0.361), (0.361,1), (1,1.22)$, the time of descent $t(\nu)$ is $\tau_2, (\tau_4,\tau_3), (\tau_B, \tau_4), >>\tau_4$. The discontinuity is at $\nu=\nu_c=0.09653$. Motion is not possible with zero speed from the point A, when $\nu>1.22.$}
\end{figure}

Though, the numerical computation underlying the result $t(0 < 0.09653)<\tau_B$ in fig. 3 have been done  using ``NIntegrate" of Mathematica that worked error-free yet we decided to re-calculate $t(\nu)$ (6) by splitting it as 
\begin{small}
\begin{multline}
t(\nu){=}\int_{0}^{\varepsilon(\nu)} F(x,y,y') dx {+} \int_{\varepsilon(\nu)}^{1} F(x,y,y') dx ,\\ \varepsilon(\nu)=10^{-n(\nu)}, n(\nu) \in N.
\end{multline}
\end{small}
The first part  in above, is very interesting where the $x$ could be extremely small in a small domain but the range of $F(x)$ consisting of large numbers. We find that for every $\nu \ge 0.0014$ one  can choose $\varepsilon(\nu)$ or $n(\nu)$ such that $I(\nu)= 3 $. For instance, when $\nu=0.01$, $n=27$. For $\nu <0.003$ $n=98$. For $\nu=.002, n=148$. For $\nu=0.0015, n=198$. For $\nu=0.0014, n= 212$. We failed to find  a suitable value of $n$ for $\nu <0.0014$ to get $I(\nu)=3$, hence $t=\tau_2 <\tau_B$. We would like to assert that $\nu =0.0014$ defines the actual critical value $\nu_c=0.0014$ below which the time of descent remains less than $\tau_B$, however the discontinuity in $t(\nu)$ in Fig. 3 has been pushed towards $\nu=0$.  One may  wonders if $I(\nu)$ 
is discontinuous at $\nu=0$ such that $I(0)=1, I(0^+)=3$  or equivalently
\begin{equation}
t(0)=\tau_2, \quad t(0^+)=3 \tau_2 > \tau_B,
\end{equation}
is expected physically.

Finally, let us resolve the curious  domain  $\nu \in (0, 0.00114)$ for the $\nu$-track (8). In this case, the track becomes practically discontinuous as
\begin{equation}
y(x=0)=1, \quad y(x=0+\varepsilon)=0,
\end{equation}
but in the energy conservation condition (1), the particle has already been assumed to have the initial potential energy equal to $mgy(0)$. Next, the integral (6) determining time of descent, inherently starts integration from  $x=0+\varepsilon$, leaving out $x=0$. Consequently, the particle executes only the linear horizontal motion from $x=0+\varepsilon$ to $x=1$ with the conditioned initial speed $v=\sqrt{2gy(0)}$ and takes time $t=\tau_2 <\tau_B$.
One may call this trivial case as a mathematical Brachistochrone (MB). In this domain of $\nu$, the deep and steep $\nu$-track would lie below even the track-{\bf f} in Fig.1. 
When the  $\nu> 0.00114$ steepness of the track reduces, the particle performs  almost vertical plus horizontal motion and we get $\tau_4<t \le  \tau_3$, so on and so forth (see Fig. 3).

The next question is how the optimization of $t$ in the earlier treatments of Brachistochrone specially in textbooks [2] have ignored this trivial yet interesting track $y(x)=0$ (13). In this regard, 
 our discussion differs slightly from textbooks as we take the starting point as (0,0) but ours  the point (0,1). So in our eq. (2), $1-y(x)$ occurs instead of $y(x)$ as in books. Optimization of $t=\int_{0}^{1} F(x,y,y') dx$ is done using  Euler-Lagrange equation, where $F=\sqrt{1+y'^2}/(1-y(x))$.  
\begin{equation}
\frac{\partial F}{\partial y} -\frac{d}{dx} \frac{\partial F}{\partial y'}=0.
\end{equation}
If we multiply (9) by $y'$ on both sides, $y'$ must be non-zero and then we get
\begin{equation}
\frac{d}{dx}\left(F-y'\frac{\partial F}{\partial y'}\right)=0 \Rightarrow F-y'\frac{\partial F}{\partial y'}=C
\end{equation}
In text books one uses (15) instead of (14),  to get the ordinary differential equation (ODE) as $(1-y)(1+y'^2)=C$, solving which in  various ways one gets the acclaimed cycloid. Hence the condition that $y'(x) \ne 0$  rules out the trajectory $y(x)=c$ from the optimization of the time of descent, without even a mention, though we can  see it as mathematically consistent case giving $t <\tau_B$.

We have discussed the time of descent of a bead/particle on various tracks/wire on several tracks from point A to the point B. We found that even convex tracks can allow the side of the bead downwards with zero-initial speed up to a limited initial  curvature of the track with $t>>\tau_3$. Concave tracks take time in the interval $(\tau_4, \tau_3)$, Between two  slightly wavy tracks the one that has initial part as concave is faster. The mathematical  Brachistochrone $y=0$ pushed by an initial speed  $v(x=0)=0=\sqrt{2g}$ m/s is  faster than even the cycloid track $\tau_2<\tau_B.$ The Numerical quadrature for $x \in (0,1)$ gives a surprising result wherein $t$ turned out to be lesser than $\tau_B$, but by splitting it in two parts (11), we could  rescue this unphysical result to some extent.  For both $\lambda$ and $\mu$ tracks, numerically $t(0^+)$ is found to be in $(\tau_4,  \tau_3$), with the warning of inaccuracy. However, for the $\nu$-track $t=\tau_2<\tau_B$ when $0<\nu <0.0014$, we believe that in this domain the track collapses to the trivial yet interesting mathematical track $y=0$ (13) and we have $t(0^+)=\tau_2$ instead of $3 \tau_2> \tau_B$. It could be challenging to develop a numerical quadrature for the integral (9) that gives $t(0^+)=3 \tau_2$.

\section*{\Large{References}}
\begin{enumerate}
	\item https://mathcurve.com/courbes2d.gb/
	brachistochrone/brachistochrone.shtml
		\item See e.g.,  Mathews and Walker, Mathematical Methods in Physics, (Pearson: New Delhi, 1979) 2nd Ed. p. 337.\\
	 H. Goldstein H, C. Poole and J. Safko, Classical Mechanics, (Pearson: New Delhi, 2002) 3rd Ed. pp. 42, 63.
	 \item G. Venezian, Am. J. Phys. {\bf34}(1966) 701.
	 \item J.M. Supplee and F. W. Schmidt, Am. J. Phys. {\bf 59} (1991) 467.
	 \item J. P Ballentine and T-P Lin,, Am. Math. Monthly {\bf 84} (1977) 652.
	 \item J Yang, D. G. Stork and David Galloway, Am. J. Phys. {\bf 55} (1987) 844.
	 \item S. G. Aiza, R. W. Gomez, W. Marquina, Eur. J. Phys. {\bf 27} (2006) 1091.
	 \item C. E.Mungan, and T. C. Lipscombe, Eur. J. Phys. {\bf 38} (2017) 1.
	 \item D. Agmon and H. Yizhaq, Eur. J. Phys. {\bf 40} (2019) 035005.
	 \item C. M. Bender, D.C. Bordy, H. F. Jones and B. K. Meister, Phys. Rev. Lett. {\bf 98} (2007) 040403.
	 \item A. Mostafazadeh, Phys. Rev. Lett. {\bf 99} (2007) 130502.
\end{enumerate} 
\end{document}